\documentclass[twocolumn,superscriptaddress,
 amsmath,amssymb,
 aps,prl]{revtex4-1}
\usepackage{graphicx}
\usepackage{bm}
\usepackage{xr-hyper}
\usepackage{hyperref}
\usepackage{xcolor}

\begin{document}

%\preprint{APS/123-QED}

\title{Optically-induced frequency up-conversion of the ferromagnetic resonance in an ultrathin garnet}

\author{Lucile Soumah}
\affiliation{Department of Physics, Stockholm University, 10691 Stockholm, Sweden}
\author{Davide Bossini}
\affiliation{Department of Physics and Center for Applied Photonics, University of Konstanz, Konstanz, Germany}
\author{Abdelmadjid Anane}
\affiliation{Unité Mixte de Physique CNRS, Thales, Univ. Paris-Sud, Université Paris Saclay, 91767, Palaiseau, France}
\author{Stefano Bonetti}
\email{stefano.bonetti@fysik.su.se}
\affiliation{Department of Physics, Stockholm University, 10691 Stockholm, Sweden}
\affiliation{Department of Molecular Sciences and Nanosystems, Ca' Foscari University of Venice, 30172 Venice, Italy}

\begin{abstract}
We perform ultrafast pump-probe measurements on a nanometer-thick crystalline Bi-doped yttrium iron garnet film with perpendicular magnetic anisotropy. Tuning the photon energy of the pump laser pulses above and below the material's bandgap, we trigger ultrafast optical and spin dynamics via both one- and two-photon absorption. Contrary to the common scenario, the optically-induced heating of the system induces an increase up to 20$\%$ of the ferromagnetic resonance frequency. We explain this unexpected result in terms of a photo-induced modification of the magnetic anisotropy, i.e. of the effective field, identifying the necessary conditions to observe this effect. Our results disclose the possibility to optically increase the magnetic eigenfrequency in nanometer-thick magnets. 
 
\end{abstract}
 
\maketitle

%\section{Introduction}

Since the discovery of the giant magnetoresistance effect \cite{baibich1988giant}, static equilibrium magnetization states are the most common method to encode digital information in data centers worldwide. To date, the dynamical properties of magnetization remains mostly unexploited in commercial technologies, with the notable exception of precise microwave frequency standards. Magnetization dynamics is attractive for technology since the frequency of magnetization precession can be tuned from the GHz to the THz range by the choice of different materials or by tuning an externally applied magnetic field \cite{gurevich1996magnetization}. Collective spin excitations -- so-called spin waves -- can also be used to write, transport or manipulate magnetic information \cite{chumak2014magnon,fischer2017experimental}. The best media to excite and propagate spin waves are materials with low magnetic damping \cite{yu2014magnetic,collet2016generation} .Yittrium iron garnet (YIG) holds the record for the lowest reported magnetic damping in bulk materials ($5\cdot10^{-5}$ \cite{yu2014magnetic}), therefore it is commonly used for spin wave devices \cite{serga2010yig}. High-quality crystalline garnet specimens with thickness of several nanometers and Gilbert damping $\alpha = 6\cdot10^{-5}$ \cite{hauser2016yttrium,soumah2018ultra,yoshimoto2018static,onbasli2014pulsed}, outperforming metallic magnetic films by an order or magnitude \cite{lee2017metallic}, can nowadays be grown. 
%Those garnet films are now used to realize complex spin-wave devices with standard lithographic techniques\cite{evelt_emission_2018,safranski_spin_2017,collet_generation_2016,collet_spin-wave_2017}.

Ultrashort light pulses have emerged as an efficient tool to initiate and probe magnetization dynamics in iron garnets on the (sub)-picosecond time-scale \cite{hansteen2005femtosecond,kirilyuk2010ultrafast,yoshimine2014phase,savochkin2017magnetization}.
The numerous studies on optically-induced magnetization dynamics in garnets have unraveled both dissipative \cite{deb2019femtosecond,shelukhin2018ultrafast} and non-dissipative \cite{,koene2016spectrally,shen2018dominant} mechanisms to trigger magnetization precession, magnetization reversal \cite{stupakiewicz2019selection} and spin wave propagation \cite{hashimoto2018frequency}. This impressive research effort has hitherto been mostly focused on relatively thick iron garnets (from 100 nm to few micrometers \cite{savochkin2017magnetization,deb2019damping,hashimoto2018frequency}), that provide large magneto-optical signals. The aforementioned experiments demonstrate that laser-induced heating in garnets is ubiquitous and it can either be the mechanism driving magnetization dynamics (change of magnetic anisotropy) or taking place simultaneously while other effects set spins in motion (opto-magnetic effects). On the other hand, ultrathin magnetic garnets have been barely studied \cite{shen2018dominant}. Despite the potential of scaling the generation and detection of coherent magnetization precession to the nanoscale, the ultrafast spin dynamics in few-nanometer-thick garnet films with perpendicular magnetic anisotropy (PMA) and low magnetic losses has not been investigated yet. We note that it follows straightforwardly from the thermodynamical description of a magnetic material and spin wave theory, that laser-induced heating can decrease the magnetic eigenfrequency, as the sample temperature increases, while an optically-induced increase of the precession frequency is not expected. However, increasing the frequency of spin dynamics in a given material without relying on a tunable external field is crucial for technological applications, since it implies a lower $1/f$ noise and an overall higher operational frequency of a possible device.   

In this work, we study light-induced magnetization dynamics on a nanometer thick Bi:YIG film with PMA. We use two different pump wavelengths to trigger magnetization dynamics with photon energies above and below the material bandgap \cite{wittekoek1975magneto,deb2012magneto}. To access both optical absorption and light-induced spin dynamics, we measure the transient transmissivity as well as the rotation of the probe polarization. The time-resolved transmissivity data show that the sample is laser-heated, via different mechanisms depending on the pump wavelength. The rotation of the polarization, that we could ascribe to the magneto-optical Faraday effect, reveal that the ferromagnetic resonance is triggered by laser-induced heating. Remarkably, we succeed in increasing the frequency of the magnetic precession up to 20$\%$ of its initial value by tuning the fluence of the photo-excitation. This unexpected observation can be understood by analysing how the absorption affects the magnetic properties of the material. 

The Bi:YIG sample studied in this work is 17 nm thick and was grown by pulsed laser deposition on a (111) oriented, substituted gadolinium gallium garnet substrate. The PMA in this sample is stabilized by both magneto-elastic and growth-induced anisotropy \cite{soumah2018ultra}. Bismuth-doping enhances the MO response compared to pristine YIG \cite{hansen1984magnetic}. The experimental set up is described in the Supplemental Material\cite{SM} . In order to investigate excitation below and above bandgap, we used two different wavelengths: 800 and 400 nm, whose corresponding photon energies (1.55 eV and 3.1 eV) are well below and slightly above the Bi:YIG bandgap at approximately 3 eV \cite{deb2012magneto,wittekoek1975magneto}. All signals are normalized with respect to the measured light intensity. The rotation of the probe polarization is additionally normalized for the different probe wavelength using the magneto-optic hysteresis loops with the magnetic field swept along the out of plane direction \cite{SM}.  We label these normalizing signals $T_0$ and $\theta_0$ for transmissivity and the probe rotation, respectively.  In order to ensure minimal cross-talking between pump and probe radiation in the photodiodes, the two beams had always different colors, either 800 or 400 nm. %Throughout the paper, orange tones in the figures indicate that the pump wavelength was at 800 nm (hence the probe was at 400 nm); conversely, blue tones in the plots show the data measured with the pump at 400 nm (hence the probe at 800 nm). 
The absolute magnitude of the optical and magneto-optical responses of the material is of course dependent on the probe wavelength. However, since we are normalizing all data at the respective wavelengths, none of the conclusions that we report is affected by this choice.

The absorption of the material is greatly increased above bandgap, and the literature reports values which differs by at least one order of magnitude between the 400 and 800 nm wavelength \cite{hansen1984magnetic,hashimoto2018frequency}.We could estimate the absorbance of our films to be approximately $5\pm1\%$ at 400 nm. At 800 nm the absorbance value was below our experimental sensitivity. Hence, in order to perform experiments in comparable \emph{absorbed} fluence regimes, we excited the film with \emph{impinging} fluence that at 800 nm pump wavelength was approximately 10 times larger than at 400 nm.

\begin{figure}[t!]
\includegraphics[width=\columnwidth]{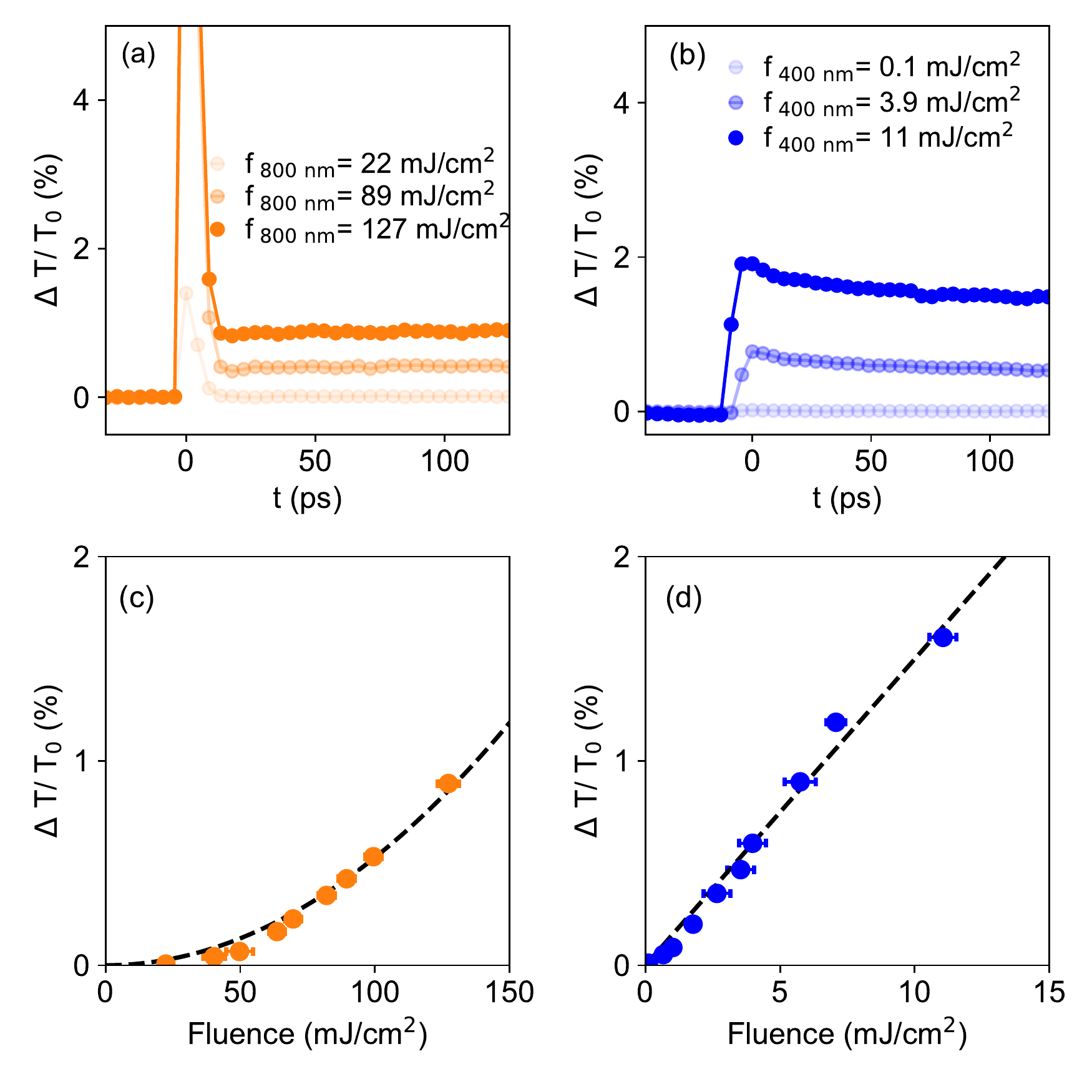}
\caption{Top panels: temporal evolution of the transmissivity of the Bi:YIG film for (a) 800 nm and (b) 400 nm pump excitation. Lower panels: fluence dependence transmissivity 100 ps after the temporal overlap for (c) 800 nm and (d) 400 nm pump excitation wavelengths. Dashed lines are fits to the curves, described in detail in the main text.}
\label{fig:fig1}
\end{figure}

The time traces of the pump-induced changes in transmissivity ($\Delta T/T_0$) for 800 nm and 400 nm pump wavelengths are displayed in Fig. \ref{fig:fig1}(a) and, \ref{fig:fig1}(b) and show qualitatively different features. For the 800 nm pump, $\Delta T/T_0$ increases sharply at the temporal overlap of pump and probe pulses, then it quickly relaxes down to the negative delays value. A substantial increase from the unpumped baseline is observed only for fluences above approximately 50 mJ/cm$^2$. For the 400 nm pump, the data do not show the same sharp increase at time zero, but it relaxes on similar times scales, i.e. hundreds of ps.

In order to get more insight on the pump absorption processes, we evaluate the change in transmissivity 100 ps after the overlap as a function of fluence. This is shown in Fig. \ref{fig:fig1}(c)-(d) for 800 nm and 400 nm respectively. For the 800 nm pump, $\Delta T/T_0$  can be well-fitted by a quadratic function. Such dependence is consistent with a two-photon absorption process, commonly reported for garnets optically excited at this wavelength \cite{hansteen2005femtosecond,koene2016spectrally,hashimoto2018frequency,deb2019femtosecond}. Conversely, the dependence of $\Delta T/T_0$ for 400 nm pump on the fluence is linear, signature of a one-photon absorption process. For a photon energy of approximately 3.1 eV, the absorption of a photon leads to direct electronic transitions of Fe$^{3+}$ ions from the ground to the excited state \cite{wittekoek1975magneto,deb2012magneto,hashimoto2018frequency}. The transmissivity traces suggest that, at both pump wavelengths, the laser pulses induce a temperature increase in the sample. %This rise is sustained over time scales longer than 900 ns, due to the insulating nature of the films and the substrate. 

\begin{figure}[t]
\includegraphics[width=\columnwidth]{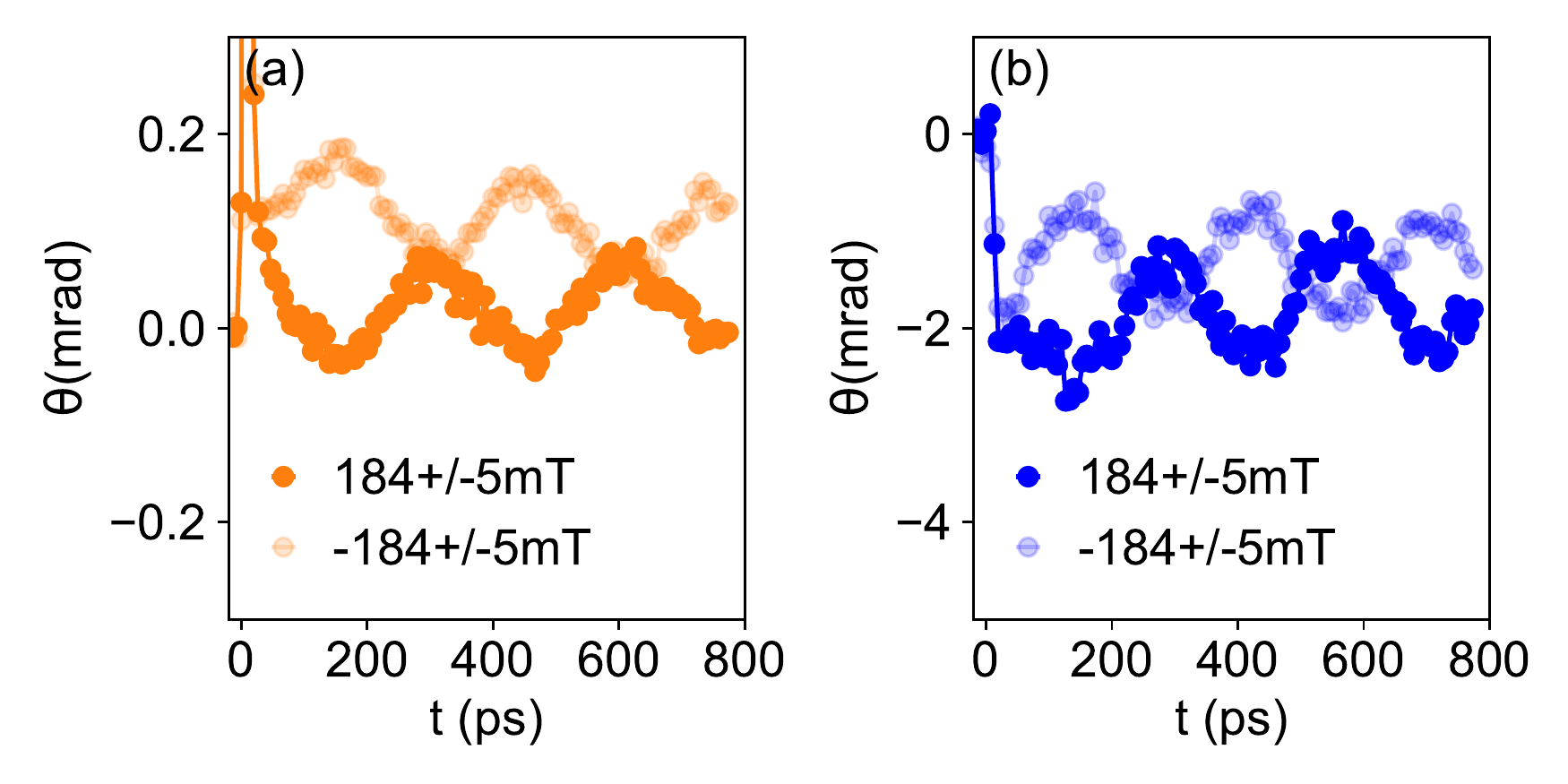}
\caption{Time-resolved Faraday rotation in the sample following the excitation using pump pulses (a) 800 nm and (b) 400 nm wavelengths, with a  fluence of approximately 10 mJ/cm$^2$ and, respectively, 0.3 mJ/cm$^2$. 
The magnitude of the externally applied magnetic field $\mu_0 H_{ext}$=184 mT was nominally the same, with its direction reverted in the semi-transparent symbols.}
\label{fig:fig2}
\end{figure}

We now turn the discussion to the time-resolved rotation of the polarization. The full response of the sample is shown in Fig. \ref{fig:fig2}. For both 800 nm and 400 nm pump wavelengths, the measured rotation shows two distinct features. First, a fast setting offset which is mostly independent of the direction of the applied magnetic field. Second, well defined and long-lived oscillations whose phase is reversed upon reversal of the externally applied magnetic field $\mu_0 H_{ext}$. This straightforward dependence on the sign of the magnetic field allows to readily disentangle the non-magnetic from the magnetic responses, by simply taking the difference between data measured with magnetic field of opposite polarities. It can be shown that the Faraday rotation (FR) is the dominating magneto-optical effect in our data.

\begin{figure}[t]
\includegraphics[width=\columnwidth]{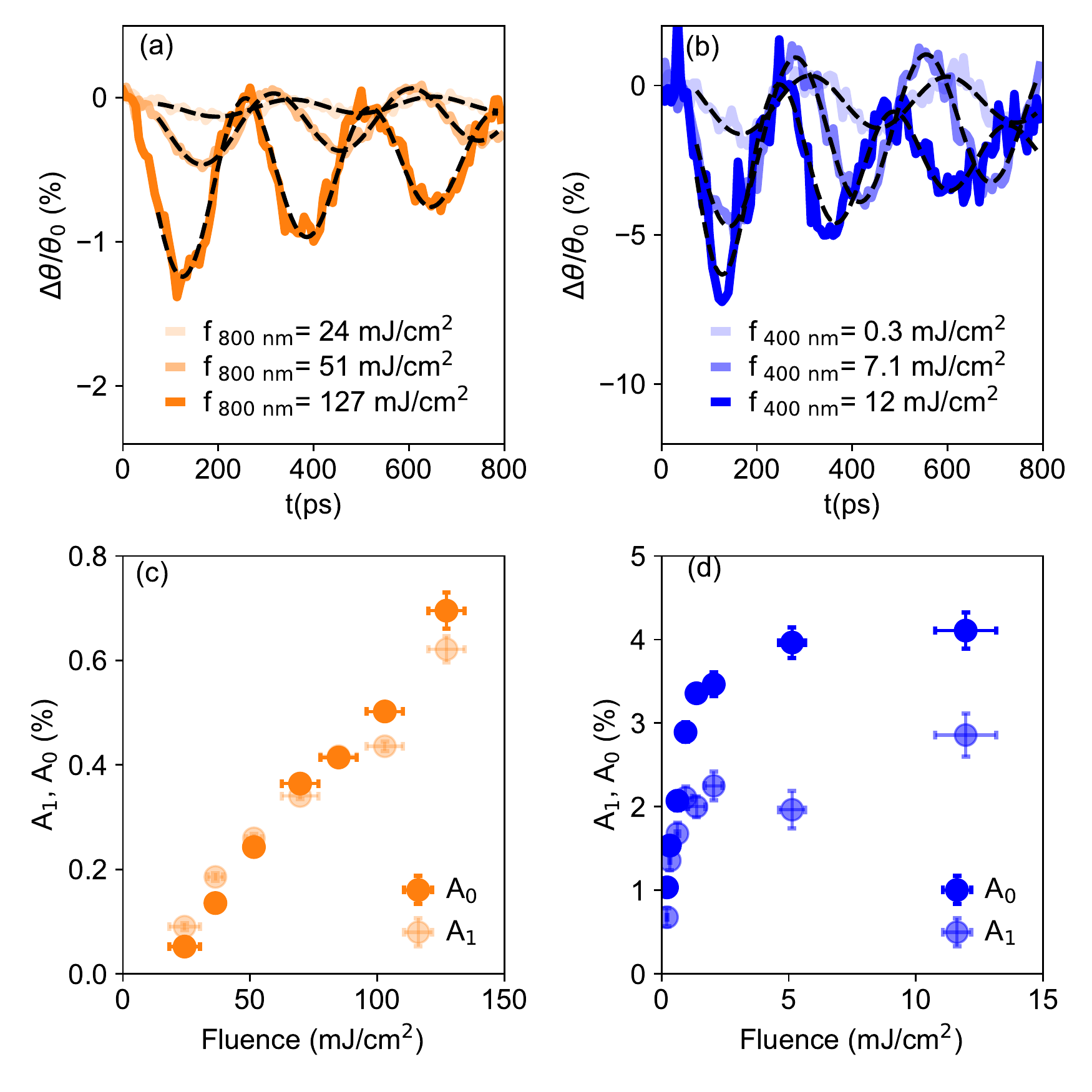}
\caption{Top panels: time-resolved Faraday rotation data for three different pump fluences at (a) 800 nm and (b) 400 nm. Dashed lines are fit to the curves using Eq. (\ref{eq:FR}). Bottom panels: fluence dependence of the fitting parameters $A_0$ (solid symbols) and $A_1$ (semi-transparent symbols) in Eq. (\ref{eq:FR}) for (c) 800 nm and (d) 400 nm pump wavelengths.}
\label{fig:fig3}
\end{figure}

To understand how the pump absorption triggers this magnetic precession we study the effect of the pump fluence on the FR signal. The difference between data at opposite field is plotted in Fig.~\ref{fig:fig3}(a)-(b) for a few selected pump fluences at both excitation wavelengths. The time traces are fitted with the expression \cite{deb2015ultrafast,koene2016spectrally,shen2018dominant} 
\begin{equation}
\theta(t)/\theta_0 = A_0 e^{-t/\tau_0}\cos \left(2\pi f_{\rm res} t + \xi \right) - A_1 e^{-t/\tau_1},
\label{eq:FR}
\end{equation}
where the first term on the right-hand-side (weighted by the factor $A_0$) describes the oscillation of the magnetization at the frequency $f_{\rm res}$, with a initial phase $\xi$, damped at the characteristic time $\tau_{0}$. The second term (scaled by $A_1$) describes all other magnetization relaxation phenomena, such as demagnetization, that can occur in the sample with a time constant $\tau_{1}$.

Fig.~\ref{fig:fig3}(c)-(d) show the extracted $A_0$ and $A_1$ parameters for the two pump wavelengths. In both cases, both terms increase monotonically with the pump fluence, although their trends are different. For the the 800 nm data, $A_0$ and $A_1$ have about the same magnitude and follow an approximately linear dependence on the fluence. For the 400 nm data, the two parameters increase up 5  mJ/cm$^2$ and then saturate. The $A_0$ parameter is approximately twice as large as $A_1$. The fit of the magnetic response using Eq. (\ref{eq:FR}) also returns the values of $f_{\rm res}$ in the range of a few GHz, which suggest that the observed oscillations are due to the ferromagnetic resonance (FMR).
In order to assess if the FMR is due to dissipative effects we measured the response of the film while varying the polarization of the pump beam at both pump wavelengths. We observed no effect of the polarization of the pump, neither on the phase nor on the amplitude of the oscillations (not shown). This strongly points to the fact that the FMR precession is heat-driven \cite{hansteen2005femtosecond,deb2015ultrafast,shelukhin2018ultrafast,deb2018picosecond}. Heat-induced FMR precession in magnetic garnets is attributed to ultrafast heating of the lattice, which causes a sudden change in the magnetic anisotropy \cite{shelukhin2018ultrafast}. In turn, this modification of the anisotropy results in a misalignment between the effective magnetic field and the magnetization, which is no longer in equilibrium. The dynamics towards the recovery of the equilibrium state is described by the Landau-Lifshitz-Gilbert equation.

\begin{figure}[t]
\includegraphics[width=\columnwidth]{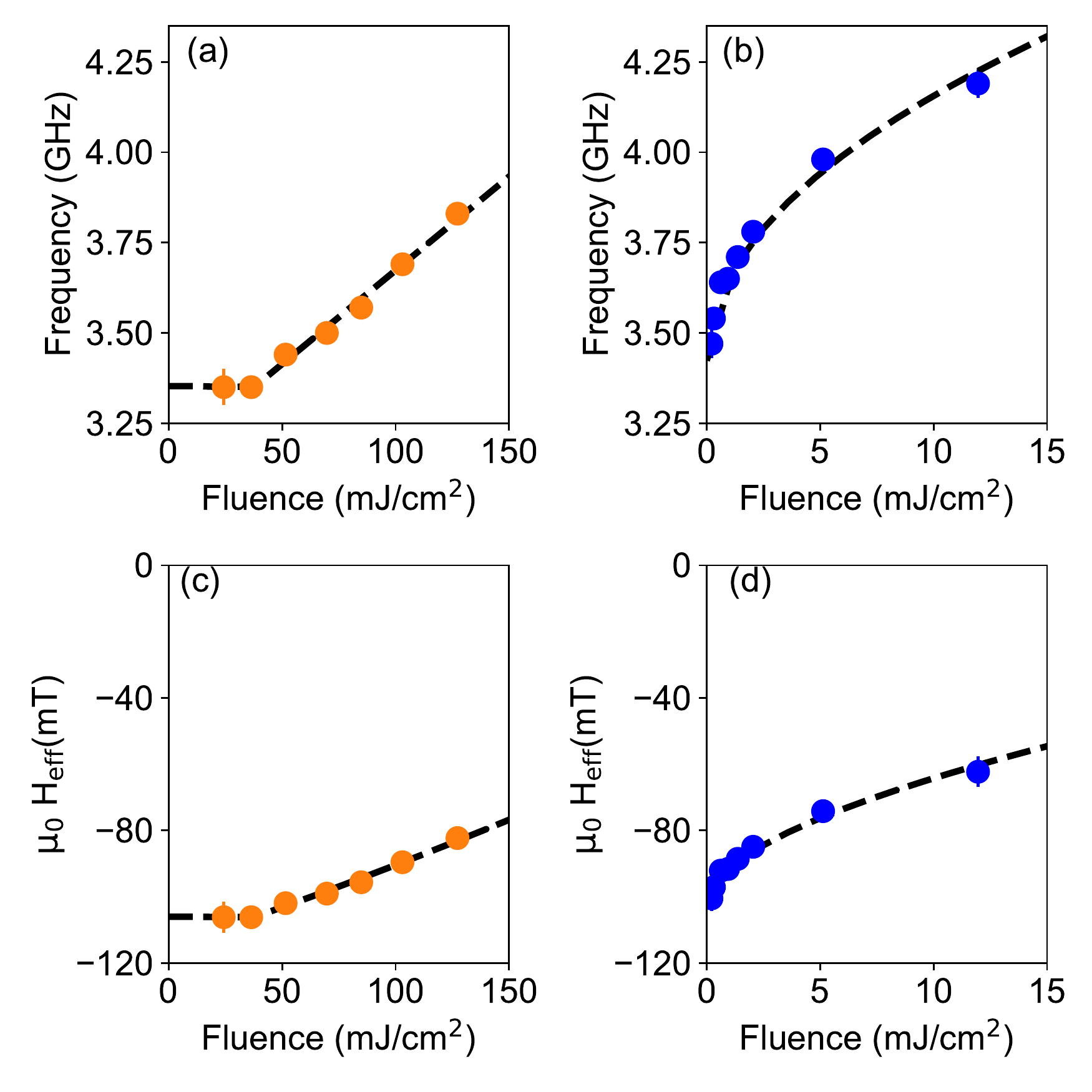}
\caption{Top panels: extracted FMR frequencies for different pump fluences at (a) 800 nm and (b) 400 nm wavelengths. Bottom panels: fluence dependence of the effective field $\mu_0 H_{\rm eff}$ calculated using Eq. (\ref{eq:heff}) for (c) 800 nm and (d) 400 nm pump wavelengths. Dashed lines in all panels are fit to the curves using the expressions given in the main text.}
\label{fig:fig4}
\end{figure}

We analyze now the mechanism inducing the FMR in detail relying on the Kittel formalism, which we assume to be valid on the relevant time-scales (tens to hundreds of ps). In this temporal regime, a dielectric can be considered to be in an equilibrium state, with an increased temperature in comparison with the state prior to the photo-excitation. from the dependence of the resonant frequency to the external field, we use the Kittel formula $f_{\rm res} = \gamma\mu_0\sqrt{H_{\rm ext}\cdot\left(H_{\rm ext}+H_{\rm eff}\right)}$ to extract the magnetic characteristics of our sample \cite{farle1998ferromagnetic}. Here, $\gamma\approx28$ GHz/T is the gyromagnetic ratio, and the effective field is $H_{\rm eff}=M_s-H_{K_U}$, with $M_s$ being the saturation magnetization and $H_{K_U}=2K_{U}/M_{s}$ the uniaxial anisotropy, with $K_U$ being the anisotropy constant. The in-plane magnetocrystalline anisotropy in this system is negligible \cite{soumah2018ultra}. The estimation of $H_{\rm eff}$ using Eq. (\ref{eq:FR}) is described in the Supplemental Material \cite{SM}.  The fit returns $\mu_0H_{\rm eff}\approx-101\pm 21$ mT. Such negative value of the effective field is consistent with a strong PMA, compensating the dipolar demagnetizing term. Furthermore, from Fig. \ref{fig:fig3}(a)-(b), it can be noticed that the frequency of the oscillations increases as the pump fluence $F$ increases. In order to get more insight in this behavior, we plot the evolution of the extracted $f_{\rm res}$ from Eq. (\ref{eq:FR}) as a function of $F$ in Fig. \ref{fig:fig4}(a)-(b) for 800 nm and 400 nm.

The increase in $f_{\rm res}$ is qualitatively different for the two cases. For the 800 nm pump, above 50 mJ/cm$^2$, the FMR resonance frequency increases linearly with $F$. At 400 nm pump, the change in resonance frequency occurs even at the lowest fluence, and it varies as the square root of $F$. The fit of the overall dependence (dashed lines) are displayed in the Fig. \ref{fig:fig4}(a)-(b), with the fitting functions being $f_{800} = 5$ GHz$\cdot$(mJ/cm$^2$)$^{-1}F+3.36$ GHz, and $f_{400} = 0.23$ GHz$\cdot$(mJ/cm$^2$)$^{-1/2}F+3.42$ GHz. The error if of the order of 1\% for all fit parameters. As we kept the direction and the magnitude of the magnetic field constant, the change in resonance frequency must be, according to the Kittel formula, attributed to a photo-induced change in the effective field $\mu_0H_{\rm eff}$ due to the optical pump. The pump induced change of $\mu_0H_{\rm eff}$ is written as
\begin{equation}
\mu_0 H_{\rm eff} = \mu_0 H_{\rm ext}-\left(\frac{f_{\rm res}}{\gamma}\right)^2\frac{1}{\mu_0 H_{\rm ext}}.
\label{eq:heff}
\end{equation}

In Fig. \ref{fig:fig4}(c)-(d) we show this change in $\mu_0H_{\rm eff}$ calculated from the Fig. \ref{fig:fig4}(a)-(b) using Eq. (\ref{eq:heff}). For both wavelengths, the magnitude of $\mu_0H_{\rm eff}$ is decreasing with fluence, suggesting that the optical pump quenches effectively the anisotropy of the film. Since the effective field includes both the magnetization and the anisotropy, we cannot exclude a concurring demagnetization of the sample. However, since the frequency is monotonously increasing, the reduction of $M_{s}$ has to be smaller than the reduction $H_{K_U}$. This is consistent with the phenomenological and theoretical relation between the temperature-dependent  magnetization  and  the uniaxial  magnetic anisotropy constant K$_U$ which according to Ref. \cite{zener1954classical} is generally written as $K_U(T)/K_U(0\,\rm K)=\left[M_s(T)/M_s(0\,\rm K)\right]^{n}$ with $n>1$.

Another evidence supporting the heat-induced modification of the effective field is provided by the different functional variations of $\mu_0H_{\rm eff}$ for the two pump wavelengths. From Eq. (\ref{eq:heff}), it can be readily inferred that the effective field $H_{\rm eff}$ is a quadratic function of the frequency. For the 800 nm pump, when the resonance frequency is a linear function, with an offset, of the fluence, the response is a second order polynomial. For small variations of the effective field, the leading term depends linearly on the fluence, and it tends towards a quadratic behavior for larger fluences. For the 400 nm pump, the resonance frequency varies as the square root of the fluence, hence the effective field has both a linear and a square root contribution, with the lower-order term dominating at smaller fluences.

The presence of the higher-order terms establishes the link between the transmissivity data of Fig. \ref{fig:fig1}, and the heat-induced spin dynamics. For the 800 nm pump, the two-photon absorption process leads to a quadratic dependence of the transmissivity with fluence, and it is likely the same process quenching the anisotropy. Similarly, the one-photon absorption explains the linear variation of the transmissivity as a function of the fluence induced by 400 nm pump, and it is the cause of the reduction of the anisotropy at this wavelength.

We briefly comment on the significance of our results. It is remarkable that optical laser pulses are able to quench the effective field of an iron garnet by up to 40\% without irreversible damages. With sample engineering, it is reasonable to expect that photo-induced full quenching of the magnetic anisotropy may be possible. We stress that the increase of the frequency is quantitatively significant (up to 20\%), finely tunable (we quantify the smallest pump induced increase of fluence to 0.07 GHz) but also counter-intuitive. Laser-induced heating is commonly excepted to decrease the frequency of magnetic eigenfrequencies, because of the temperature increase. Strikingly, the higher the fluence of the pump beam, the stronger the increase of the frequency, although simultaneously the temperature increase is also enhanced. In our case, due to the significant contribution of the uniaxial anisotropy term in the effective field, we observe the opposite behaviour.

We are not aware of other material systems where a similar phenomenon has been achieved.  We observe that although nonlinear spin dynamics has already been reported in a limited variety of dielectric magnetic materials, the mechanisms are different than the concept discussed here. In particular, rather different mechanisms such as magneto-acoustic coupling \cite{afanasiev2014laser} and strong THz resonant \cite{baierl2016terahertz,mukai2016nonlinear} or non-resonant \cite{hudl2019nonlinear} pumping were exploited to induce harmonics of the spin resonances. Hence, our findings reveal a novel mechanism of anisotropy control that can only be achieved in ultrathin iron garnets with PMA. Given that these garnets also possess an extremely low magnetic damping, our results pave the way for new control schemes in spin wave devices where their operating frequencies can be arbitrarily controlled by light.

%\section{Conclusion}
%In conclusion, we performed optical pump-probe measurements on a 17 nm thick Bi:YIG films with perpendicular magnetic anisotropy. We used two different wavelengths, above and below the bandgap of the material, to measure the time-resolved transmissivity and Faraday rotation. Both quantities showed qualitatively different fluence dependencies for the two wavelengths, consistent, however, with a common heat-driven excitation mechanism of the sample. The Faraday rotation data showed that ultrafast laser pulses can be efficiently used to trigger magnetization dynamics in our sample. In addition, the FMR frequency is tuned by more than 20\% simply by varying the pump fluence, which has the effect of  the perpendicular magnetic anisotropy. 
L.S.\ and S.B.\ acknowledge support from the Knut and Alice Wallenberg Foundation, grant number 2017.0158. D.B acknowledges support from the Deutsche Forschungsgemeinschaft (DFG) program BO 5074/1-1.

\end{document}